\documentclass[prl,aps,twocolumn,amsfonts,showpacs,footinbib,preprintnumbers,superscriptaddress]{revtex4-1}

\usepackage{amsmath,amsthm,amssymb} 
\usepackage{dsfont,yfonts}
\usepackage{hyperref}
\usepackage{floatrow,dcolumn,rotating}
\usepackage{array,xcolor,graphicx}

\hypersetup{colorlinks=true,linkcolor=blue,citecolor=red,urlcolor=blue}

\newcommand{\nn}{\nonumber\\}

\def\CL{\mathcal{L}}

\def\CN{\mathcal{N}}
\def\CO{\mathcal{O}}

\def\CV{\mathcal{V}}

\def\qfr{\mathfrak{q}}
\def\wfr{\mathfrak{w}}
\def\lgb{\lambda_{\scriptscriptstyle GB}}
\def\ngb{N_{\scriptscriptstyle GB}}

\begin{document}
\preprint{MIT-CTP/4940}

\title{Black hole scrambling from hydrodynamics}

\author{Sa\v{s}o Grozdanov}
\affiliation{Center for Theoretical Physics, MIT, Cambridge, MA 02139, USA}
\affiliation{Instituut-Lorentz for Theoretical Physics $\Delta$ITP,
  Leiden University, Niels Bohrweg 2, Leiden 2333 CA, The Netherlands}
\author{Koenraad Schalm}
\affiliation{Instituut-Lorentz for Theoretical Physics $\Delta$ITP,
  Leiden University, Niels Bohrweg 2, Leiden 2333 CA, The Netherlands}
\author{Vincenzo Scopelliti}
\affiliation{Instituut-Lorentz for Theoretical Physics $\Delta$ITP, Leiden University, Niels Bohrweg 2, Leiden 2333 CA, The Netherlands}

\begin{abstract} 
We argue that the gravitational shock wave computation used to extract the scrambling rate in strongly coupled quantum theories with a holographic dual is directly related to probing the system's hydrodynamic sound modes. The information recovered from the shock wave can be reconstructed in terms of purely diffusion-like, linearized gravitational waves at the horizon of a single-sided black hole with specific regularity-enforced imaginary values of frequency and momentum. In two-derivative bulk theories, this horizon ``diffusion'' can be related to late-time momentum diffusion via a simple relation, which ceases to hold in higher-derivative theories. We then show that the same values of imaginary frequency and momentum follow from a dispersion relation of a hydrodynamic sound mode. The frequency, momentum and group velocity give the holographic Lyapunov exponent and the butterfly velocity. Moreover, at this special point along the sound dispersion relation curve, the residue of the retarded longitudinal stress-energy tensor two-point function vanishes. This establishes a direct link between a hydrodynamic sound mode at an analytically continued, imaginary momentum and the holographic butterfly effect. Furthermore, our results imply that infinitely strongly coupled, large-$N_c$ holographic theories exhibit properties similar to classical dilute gasses; there, late-time equilibration and early-time scrambling are also controlled by the same dynamics.
\end{abstract}

\maketitle

{\bf Introduction.---}The notion that dynamics at widely separated
timescales is governed by independent processes lies at the heart of
modern physics. The emergence of collective phenomena is a clear
example. At very short timescales, the physics is described by
microscopic ``far-from-equilibrium'' dynamics; at long timescales, it
is the universal statistics-dominated processes that control the onset
of equilibrium. Ironically, the most prevalent textbook example of
collective emergence, the computation by Maxwell of the shear
viscosity of a classical ideal gas, fails this guideline. As is well
known, in dilute gases the shear viscosity and some other
transport coefficients are directly related to the 2-to-2 scattering
rates of the microscopic constituents. In dilute gases, the early-time
physics thus also controls the late-time approach to equilibrium. Our
full understanding of kinetic theory explains why dilute
gases violate the canonical notion of separation of scales. The dilute gas is a special case for which the BBGKY hierarchy that builds up the long-time behavior from microscopic processes truncates \cite{chapman-book,kvasnikov-book,saint-raymond-book,ferziger-kaper-book,ford-book,groot-book,silin-book,grad-1963,gross-1959}. 

On the other hand, in generic (e.g. dense) many-body systems, the
early-time physics is distinct from late-time evolution. Of course,
this does not imply that the early-time physics is irrelevant to
collective behavior, as indeed, it crucially ensures ergodicity or
mixing (scrambling). Nevertheless, one generically distinguishes (at
least) two timescales: an early-time ergodic and a late-time
collective scale. In classical systems, ergodicity is driven by
chaotic non-linear dynamics, whereas statistics and universality
drive collective behavior. These two different scales have a direct manifestation in classical
dynamical systems analysis. Chaotic dynamics is characterized by Lyapunov exponents encoding the exponential divergence of trajectories with infinitesimally different initial conditions---the butterfly
effect. A Gibbs ensemble of such initial conditions,
however, equilibrates with a generically distinct characteristic
timescale set by Pollicott-Ruelle resonances
\cite{Polchinski:2015cea,pollicott1985rate,ruelle1986resonances},
again exemplifying the notion that widely separated timescales are
driven by different physics. 

Perturbative quantum field theories are usually studied in the dilute
regime and as in the classical gas, both timescales are driven by the
same physics \cite{Stanford:2015owe,Aleiner:2016eni,Patel:2016wdy,Chowdhury:2017jzb,Patel:2017vfp,Grozdanov:2017forthcoming}. Strongly coupled, dense, quantum theories on the other hand are
expected to have distinct scales. Triggered by studies
\cite{Hayden:2007cs,Sekino:2008he,Shenker:2013pqa,Maldacena:2015waa,Polchinski:2015cea}
on collective dynamics in strongly coupled large-$N_c$
quantum systems holographically dual to black holes, Blake observed that in the simplest such systems, late-time diffusion and early-time ergodic dynamics do appear to be governed by the same physics \cite{Blake:2016wvh}, similar to the dilute gas rather than the generic expectation. Follow-up studies extended the range of systems \cite{Blake:2016sud,Blake:2016jnn,Ling:2016ibq,Gu:2016oyy,Kolekar:2016yzg}, found counterexamples \cite{Lucas:2016yfl} and observed that it only applied to thermal diffusivity \cite{Davison:2016ngz,Blake:2017qgd}. 

In this work, we will show how the holographic computations of
quantum ergodic dynamics---the holographic butterfly effect---and
hydrodynamics are related. In particular, we will show that the
characteristic exponential growth exists on the level of (retarded) two-point functions when the
hydrodynamic sound mode is driven to instability by a choice of a
specific value of momentum. This result indicates an intriguing similarity between the behavior of
infinitely strongly coupled large-$N_c$ theories holographically dual
to two-derivative gravity and classical dilute gases in the sense that
chaotic dynamics is entirely describable by the same physics of
hydrodynamic modes, albeit excited outside of the hydrodynamic regime
of small frequency $\omega$ and momentum $k$ compared to the temperature scale $T$ of the CFT. In this Letter, we will only focus on charge-neutral systems, although we expect our findings to be valid also for charged states and for systems with momentum relaxation in which long-lived longitudinal modes are controlled by diffusion. 
  
{\bf Scrambling and hydrodynamical transport.---}By convention, the early-time onset of ergodicity is characterized by the scrambling rate $\lambda$ and the butterfly velocity $v_B$, which are defined from the early-time rate of exponential growth of 
out-of-time-ordered correlation function (OTOC) of local (unbounded) operators,
\begin{align}
C(t, x) &= - \frac{\langle
  [\hat{W}(t,x),\hat{V}(0)]^{\dagger}[\hat{W}(t,x),\hat{V}(0)]\rangle_{\beta}}{2\langle\hat{W}(t,x)\hat{W}(t,x)\rangle_{\beta} \langle \hat{V}(0)\hat{V}(0)\rangle_{\beta} }\nn
& \simeq e^{2\lambda (t - x/ v_B)} \, . \label{eq:1}
\end{align}
Here, $\hat{V}(t,x)$ and $\hat{W}(t,x)$ are generic operators, and expectation values are taken in the thermal ensemble with temperature $T=1/\beta$.
In systems with a classical analogue for which such growth persists as $t\to\infty$ and for special (unbounded) operators, this indeed computes the Lyapunov exponent $\lambda_L = \lambda$ associated with chaotic behavior underpinning classical ergodicity \cite{1969JETP...28.1200L,1996PhRvB..5414423A,1997PhRvE..55.1243A}. 

Not all systems exhibit late-time regime of exponential growth of this
correlator---in fact, most quantum systems do not
\cite{Roberts:2016wdl,Kukuljan:2017xag}, illustrating the tension
between classical chaos and quantum dynamics. Large-$N_c$ systems with a
holographic dual do exhibit such growth. Extrapolating from the
  insight that any perturbation carries energy, it has been argued
that this exponential rate can be read off from a gravitational shock
wave propagating along the double-sided (maximally extended) black
hole horizon \cite{Shenker:2013pqa}. 

The non-linear shock wave calculation implicitly focusses on
energy-momentum dynamics in the dual theory, rather than generic
dynamics, as this is what purely gravitational spacetime dynamics and
waves encode. On the other hand, the collective late-time dynamics of energy-momentum is
also well-understood with its IR dynamics governed by
hydrodynamics. Its behavior can be computed from linearized
gravitational perturbations (see e.g. \cite{Policastro:2002se,Kovtun:2005ev,Kovtun:2012rj}). The mere
fact that the gravitational shock wave encoding early-time ergodicity
describes the dynamics of energy-momentum, as do hydrodynamic
excitations, is far from sufficient for establishing any relation
between them. A more telling fact is that the exact non-linear shock
wave solution is actually also a solution to linearized gravitational
equations. This is what we show now. This results then leads to our
discovery that when perturbed with a special imaginary momentum, the
late-time hydrodynamic sound mode reflects the
leading-order early-time instability of the system with the
exponential growth set by $\lambda_L$ and the butterfly velocity $v_B$.  

{\bf Shock waves from linearized gravitational perturbations.---}Chaotic properties normally extracted from shock waves can be inferred directly from a single-sided, linearized analysis of the bulk gravitational equations.  We study five-dimensional, two-derivative, classical gravity with the action
\begin{align}
  \label{eq:6}
  S = \frac{1}{2\kappa_5^2}\int\! d^5x \sqrt{-g}\left[R + \frac{12}{L^2}
  +{\cal L}_{\scriptsize{matter}}\right] ,
\end{align}
which gives rise to the following Einstein's equations (in units where $L=1$):
\begin{align}\label{EoM}
\mathbf{G}_{\mu\nu} \equiv R_{\mu\nu} - \frac{1}{2} g_{\mu\nu} R - 6 g_{\mu\nu} = \kappa_5^2 \, T^{\scriptsize{matter}}_{\mu\nu} \,.
\end{align}
In the longitudinal sound channel, in the $h_{\mu z} = 0$ gauge with momentum in the $z$-direction, we write a first-order perturbed metric as
\begin{align}\label{Metric}
&ds^2 = - f(r) dt^2 + \frac{dr^2}{f(r)} + b(r) \left(dx^2 +dy^2 +dz^2\right) \\
&- \left[ f(r) H_1  dt^2 - 2 H_2  dt dr + \frac{ H_3  dr^2}{f(r)}  + H_4 \left(dx^2 +dy^2 \right) \right] ,\nonumber
\end{align}
where $H_i$ are functions of $t$, $z$ and $r$, and $f(r_h) = 0$. We demand that the perturbation is null in the radial direction at the horizon, set $H_4 (r_h) = 0$ and write
\begin{align}
H_1 = H_3 &= \left(C_+ W_+ (t,z,r) + C_- W_- (t,z,r) \right) , \\
H_2&=\left(C_+ W_+ (t,z,r) - C_- W_- (t,z,r) \right) .
\end{align}
First, consider $T^{matter}_{\mu\nu} = 0$ to focus on the AdS-Schwarzschild black brane background with $b(r) = r^2$ dual to thermal $\CN=4$ supersymmetric Yang-Mills (SYM) theory. We can write $W_\pm$ as
\begin{align}
W_{\pm} (t,z,r) &= e^{ - i \omega \left[ t \pm \int^r \frac{dr'}{f(r')} \right]  + i k z } \, h_\pm (r) \,, 
\end{align}
where $h_\pm (r)$ are regular at $r=r_h$. 
Using $\mathbf{G}_{rr} = 0$, then
\begin{align}\label{hsol}
h_\pm (r) &= e^{ \int^r \frac{k^2 \pm 9 i \omega r' - 12 r'^2}{3 r' f(r')} dr' } \,.
\end{align}
Imposing regularity \cite{Son:2002sd,Herzog:2002pc} on \eqref{hsol} fixes a single relation between $\omega$, $k^2$ and $r_h$. Ensuring the remaining equations of motion \eqref{EoM} are solved at $r=r_h$, gives a second, (advanced and retarded) {\em diffusive} condition,
\begin{align}
\omega_\pm \equiv \pm \, i \mathfrak{D} k^2 = \pm \,i \frac{1}{3\pi T} k^2 \,.
\end{align}
Combined with the horizon-regularity, this fixes the solution in terms of a specific imaginary momentum mode
\begin{align}
k^2 &\equiv - \mu^2  = - 6 \pi^2 T^2  \,, \label{Momentum}
\end{align}
which gives the Lyapunov exponent and the butterfly velocity, i.e. for modes with $e^{-i\omega t + i k z}$,  
\begin{align}
\omega_\pm &\equiv \mp i \lambda_L \,, ~~~ \lambda_L = 2\pi T\,, \label{LyapunovExponent} \\
v_B &\equiv \left| \frac{ \omega_\pm}{k} \right| = \sqrt{\lambda_L\mathfrak{D}} \,.
\end{align}
Away from the horizon, the corrections to the present solution can be consistently constructed in a small $\sqrt{r-r_h}$ expansion, requiring $H_4 \neq 0$.

For a regular $T_{\mu\nu}^{matter}\neq 0$, one can see the horizon diffusion arise more generally. For a background metric \eqref{Metric}, the regularity of $\mathbf{G}_{rr}$ implies
\begin{align}\label{GenBacId}
b(r_h) = b'(r_h) f'(r_h)  / 8 \, .
\end{align}
Assuming that $T_{tr} (r_h) = 0$, it follows immediately from $\mathbf{G}_{tr} (r_h) = 0$ that at $r=r_h$,
\begin{align}
\partial_t W_\pm = \mp \, \mathfrak{D} \, \partial^2_z W_\pm \,,
\end{align}
with the horizon diffusion coefficient, as in \cite{Blake:2016wvh}:
\begin{align}\label{HorizonDiffusion}
\mathfrak{D} = \frac{v_B^2}{\lambda_L} = \frac{2}{3} \frac{1}{  b'(r_h)}= \frac{1}{12}\frac{f'(r_h)}{b(r_h)}  \, .
\end{align}
Assuming that the solution is not supported by $T_{\mu\nu}^{matter}$ and requiring regularity in $\mathbf{G}_{rr}$, we again obtain the Lyapunov exponent from Eq. \eqref{LyapunovExponent} and imaginary momentum
\begin{align}
k^2 = - \frac{3}{4} b'(r_h) f'(r_h)  = - 3 \pi T \, b'(r_h)\,.
\end{align}  
Therefore, we have recovered all known shock wave results from a linear gravitational perturbation of a single-sided black brane. The validity of this solution requires sufficient decoupling of $\CL_{matter}$ at the horizon, which is implicitly assumed in the shock wave computation. Generically, this will not be the case. The sound channel couples all scalar excitations, and one needs to demand that all their equations of motion are satisfied as well.

Higher-derivative gravity corrections encode (inverse) coupling constant corrections in the dual field theory \cite{Gubser:1998nz,Kats:2007mq,Brigante:2007nu,Grozdanov:2014kva,Stricker:2013lma,Waeber:2015oka,Grozdanov:2016vgg,Grozdanov:2016zjj,Grozdanov:2016fkt}. An analogous calculation as in two-derivative theories can now be done e.g. in Gauss-Bonnet theory (for details regarding the theory see e.g. \cite{Grozdanov:2016fkt}), where we also recover the known results of Ref. \cite{Roberts:2014isa}, \footnote{$\ngb$ is conventionally set to $\ngb^2 = \left(1 + \sqrt{1-4\lgb} \right) / 2$, which ensures that the boundary speed of light is one. }
\begin{align}
\omega_\pm = \mp 2 i \pi T \, , && k^2 = - \frac{6\pi^2T^2}{\ngb^2} \,,&& v_B^2 = \frac{2}{3} \ngb^2\,.
\end{align}

Focusing again on the two-derivative action \eqref{eq:6} dual to $\CN = 4$ SYM at large $N_c$ and infinite coupling, and transforming the metric \eqref{Metric} to Kruskal-Szekeres coordinates, one finds
\begin{align}\label{WavesMetricKS}
ds^2 = &\, A(UV) \, dU dV  + B(UV) d\vec x^2 \nn
&- A(UV) \, e^{ikz} \left( C_+ \frac{dU^2}{U} -  C_- \frac{dV^2}{V} \right) .
\end{align}
Our solution thus takes the form of the exact shock wave solution $ds^2 =  A(UV) dU dV  + B(UV) d\vec x^2 - A(UV) \delta(U) h(x) dU^2$, but travelling along both null \mbox{$U=0$} and $V=0$. The only difference is that the shock solution has a Dirac delta function support $h_{UU} \propto \delta(U)$, whereas the solution presented here has support given by a (smeared) $h_{UU} \propto \Delta(U) \equiv 1 / U$. At the level of the linearized Einstein's equations, the function $\Delta(U)\equiv 1/U$ satisfies the distributional identities used to construct the shock wave solution: $U \partial_U \Delta(U) = - \Delta(U)$ and $U^2 \partial_U^2 \Delta(U) = 2 \Delta(U)$. Distributional identities of the type $F(U)\Delta^2(U) \approx 0$, when integrated over $U$ for sufficiently smooth $F(U)$, are satisfied approximately but not exactly as with $\delta^2(U)$ (see e.g. \cite{Sfetsos:1994xa}). A distinct difference is that the $\delta(U)$-shock is supported by energy-momentum at the horizon. The linearized solution \eqref{WavesMetricKS} with a less singular support is a leading-order in $1/U$ approximation of an exact smooth solution to Einstein's equation with no source of energy-momentum. It is a longitudinal (sound) mode, which encodes the correct Lyapunov exponent and the butterfly velocity.

{\bf Hydrodynamics and the sound mode.---}Sound is well understood as a hydrodynamical phenomenon. In holography, it is encoded by the low-energy limit of the sound channel spectrum \cite{Policastro:2002tn,Kovtun:2005ev} and is described by a pair of longest-lived modes $\omega^*_\pm(k)$. Within the hydrodynamic approximation (expansion of $\omega^*_\pm$ for $|k| / T \ll 1$),
\begin{align}\label{soundmode}
\omega^*_\pm(k) \approx \pm \sum_{n=0}^\infty \CV_{2n+1} k^{2n+1} - i  \sum_{n=0}^\infty \Gamma_{2n+2} k^{2n+2} ,\,
\end{align}
which is analytically known for $\CN = 4$ SYM to $\CO(k^4)$ at
infinite coupling, i.e. to third order in the hydrodynamic expansion
\cite{Grozdanov:2015kqa}. All $\CV_{n}$ and $\Gamma_n$ are real and
for $\CN = 4$ SYM at infinite coupling, $\CV_1 = 1/\sqrt{3}$, $\Gamma_2 = 1/(6\pi T)$,
$\CV_3 = (3-2\ln 2)/(24\sqrt{3}\pi^2 T^2)$ and $\Gamma_4 = (\pi^2 - 24
+ 24\ln 2 - 12 \ln^2 2) / (864 \pi^3 T^3)$. For real $k$,
Eq. \eqref{soundmode} describes attenuated propagating modes. However,
for imaginary $k$, which is required to construct the above
gravitational solution, both $\omega^*_\pm$ and $k$ are purely
imaginary. To find $\omega^*_\pm(k)$ for imaginary $k$, we compute the quasinormal mode
spectrum (poles of the retarded sound channel stress-energy tensor two-point function,  e.g. the energy-energy $G^R_{T^{00}T^{00}} (\omega, k)$) \cite{Kovtun:2005ev}, which can be done analytically in the
hydrodynamic expansion (small $|\omega|/T \ll 1 $, $|k| / T \ll 1$) or
numerically in the holographic model for any $\omega$ and
$k$. Our first observation is that for imaginary $k$, the system
is driven to instability, which results in at least one of the two
sound modes in \eqref{soundmode} having $\text{Im} [ \omega ] > 0
$. Our main result, however, is that the fully numerically computed frequency (dispersion relation) of the most unstable sound mode $\omega^*_+$ asymptotically approaches the Lyapunov exponent growth rate $k = i \mu = \sqrt{6} i \pi T$:
\begin{align}  
\omega^*_+ (i \mu) = i \lambda_L \,.
\end{align}
Precisely at $k = i \mu$, the quasinormal mode solution does not exist, even though it exists infinitesimally close to this point when approached from either side along the imaginary $k$ dispersion curve. This allows us to deduce that at the special point $\omega^*_+(i\mu)$, cf. Eqs. \eqref{Momentum} and \eqref{LyapunovExponent}, the retarded longitudinal two-point function of the stress-energy tensor has a hydrodynamic pole which contains all information about many-body chaos, $\lambda_L$ and $v_B$. Furthermore, at the point of chaos, its residue vanishes: 
\begin{align}\label{Residue}
\text{Res}  \, G^R_{T^{00}T^{00}} (\omega = \omega^*_+ (i \mu)= i \lambda_L, k=i\mu) = 0 \,.
\end{align}  
The two-point correlator identity \eqref{Residue} is sufficient for uniquely specifying the point of chaos in the CFT, eliminating the need for the OTOC considerations to find $\mu$.

\begin{figure}[h]
\centering
\includegraphics[width=1\textwidth]{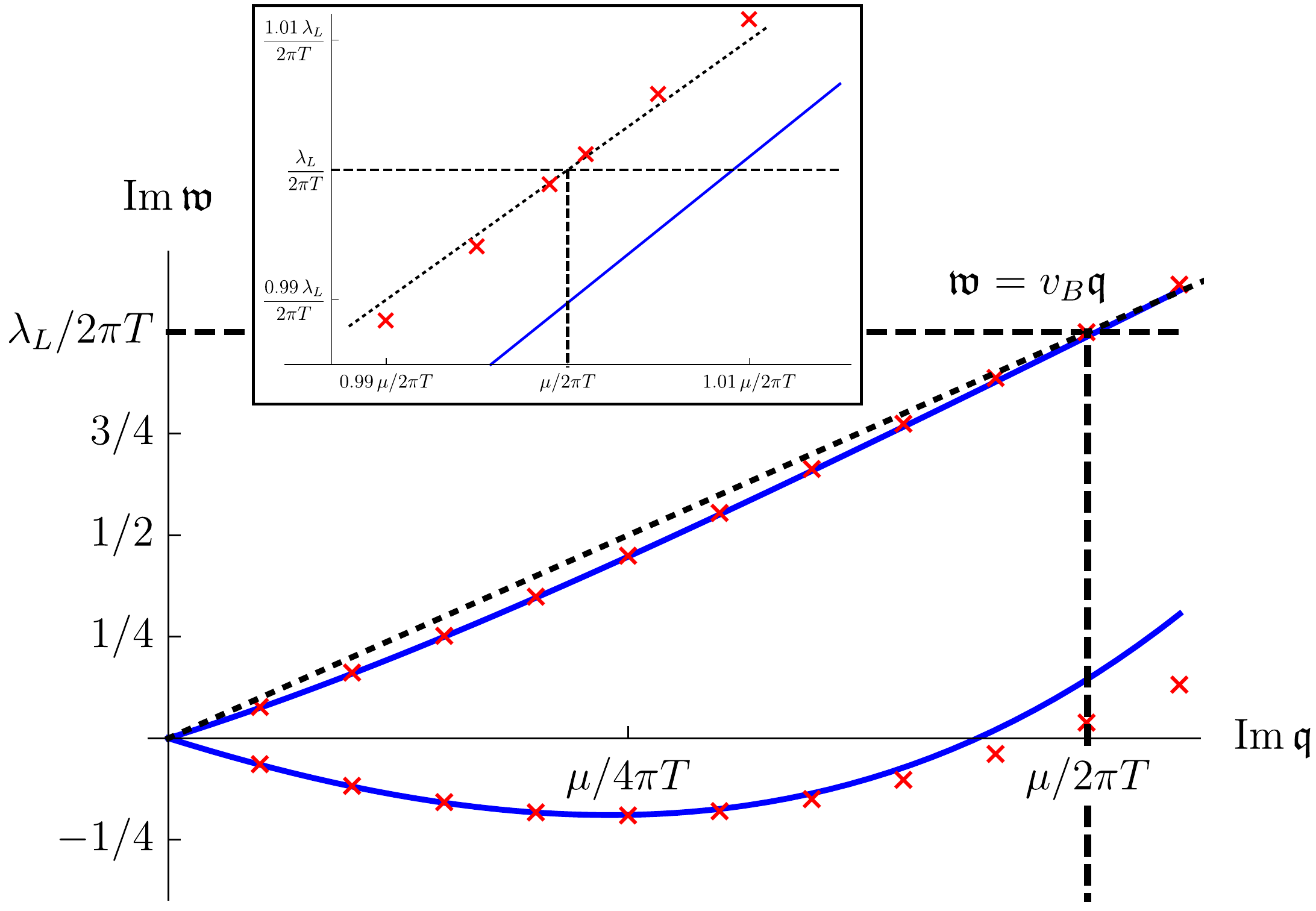}
\caption{Dispersion relations of the hydrodynamic sound modes, plotted for imaginary dimensionless $\wfr \equiv \omega / 2 \pi T$ and $\qfr \equiv  k / 2\pi T$. The blue lines depict the third-order hydrodynamic result \cite{Grozdanov:2015kqa} and the red crosses the numerically computed $\wfr^*_\pm (\qfr)$. Dashed lines indicate the values of $\omega = i \lambda_L$ and $k = i \mu$. The dotted line is the linear dispersion relation $\wfr = v_B \qfr$. The inlay depicts a zoomed-in plot around $k = i \mu$.}
\label{fig:qnms}
\end{figure}

We note that, intriguingly, the dispersion relation around this point can be reasonably well approximated by $\omega = v_B k$. This is evident from the numerical computations and from the third-order hydrodynamic approximation to $\omega_+(k)$, which reproduces the full dispersion relation of the dominant mode rather well, giving $\omega^*_+(i\mu) \approx 0.990\times i \lambda_L$. Our results are presented in Fig.~\ref{fig:qnms}.

{\bf Discussion.---}These results show that the holographic butterfly effect and black hole scrambling can be understood in terms of a hydrodynamic sound mode at a specific imaginary momentum (exponentially spatially growing fluid profile), which is fixed by dual Einstein's equations governing a radially null sound mode and the condition of regularity (without additional energy-momentum) at the horizon. At $\omega^*_+(i\mu)$, the sound mode dispersion relation gives the Lyapunov exponent associated with holographic many-body chaos. Furthermore, even though $|k|/T$ lies at the edge or outside of the hydrodynamic regime \cite{Chesler:2010bi,Grozdanov:2016vgg}, the full dispersion relation is well described by the hydrodynamic approximation.   

What are the physical implications of our observations? Several recent papers have speculated on relations between late-time diffusion and the butterfly effect \cite{Blake:2017qgd,Blake:2016jnn,Blake:2016sud,Blake:2016wvh,Lucas:2016yfl,Patel:2017vfp,Chowdhury:2017jzb,Gu:2017ohj,Hartman:2017hhp}. The late-time behavior of hydrodynamic excitations in a translationally invariant, uncharged CFTs is controlled by momentum diffusion. In theories holographically dual to two-derivative gravity, momentum diffusion $D$ is completely determined by horizon data \cite{Kovtun:2003wp} while charge diffusion is not. Given a (background) metric \eqref{Metric} and the shock wave diffusivity $\mathfrak{D}$ (cf. Eq. \eqref{HorizonDiffusion}):
\begin{align}\label{DiffRatio}
\frac{D}{\mathfrak{D}} = \frac{3 \, b'(r_h)}{8 \pi T } \,.
\end{align}
In large-$N_c$ $\CN=4$ SYM theory at infinite coupling, this reduces to $D / \mathfrak{D} = 3/4$. However, as we move away from infinite coupling and consider higher-derivative bulk theories, $D$ is no longer computable in terms of simple horizon data, which results in deviations of $\eta/ s$ from $1/4\pi$. Since the butterfly velocity and the Lyapunov
exponent are by construction computed only at the horizon, we have a-priori no reason to expect that there continues to exist a simple relation between $D$ and $\mathfrak{D}$ in holographic duals with more
then two derivatives. Indeed, the ratio of $D / \mathfrak{D}$ in Gauss-Bonnet has non-trivial coupling dependence \cite{Grozdanov:2016fkt,Roberts:2014isa}, and is thus not universal \footnote{For a discussion regarding the validity of hydrodynamics in the presence of coupling constant corrections, see \cite{Romatschke:2015gic,Grozdanov:2016vgg,Grozdanov:2016fkt,Grozdanov:2016zjj,vanderSchee:2017edo,DiNunno:2017obv}.}. 

As we emphasized in the Introduction, a relation, such as \eqref{DiffRatio}, which depends only on $r_h \sim T$, between late-time and early-time physics is rather unexpected. The exception is the classical dilute gas. Its early-time chaos and late-time diffusion are controlled by the same process (2-to-2 scattering). Our findings show that the situation is similar in an infinitely strongly coupled, large-$N_c$ CFT. As a result, early-time scrambling and late-time hydrodynamics are qualitatively related and appear to be driven by the same physics---hydrodynamics.

The reason that an obtuse relation between microscopic ergodicity from shock waves and late-time diffusion is sought after is that black holes are special in that their ergodicity rate $\lambda_L$ saturates a conjectured bound $\lambda_L \leq 2\pi T$ \cite{Maldacena:2015waa}. If early-time ergodicity indeed controlled late-time diffusion, this bound could imply a long-sought fundamental diffusion bound \cite{Hartnoll:2014lpa,Hartman:2017hhp} \footnote{For two examples of rigorous diffusion bounds in one-dimensional systems, see \cite{PhysRevE.89.012142}. In holography, one can derive bounds on conductivities in disordered systems \cite{Grozdanov:2015qia,Grozdanov:2015djs} but as of yet, not on diffusion.}. Such a fundamental bound was re-postulated several years ago based on early results on collective dynamics in holography by noting that the shear viscosity in these systems only depends on horizon data \cite{Policastro:2001yc,Kovtun:2004de}. Expressions such as Eq. \eqref{DiffRatio} make it clear, however, that
the Lyapunov exponent bound does not yield a diffusion bound. The
dependence on the temperature through $r_h$ or the presence of
additional scales allows this ratio to take any value. We note that
such temperature dependence is also present in the classical dilute
gas of particles with mass $m$ and density $\rho$ through the average
velocity $v$. Its shear viscosity $\eta$ and the Lyapunov exponent \cite{PhysRevLett.80.2035} behave as 
\begin{align}\label{eq:5}
\eta \sim m\frac{\sqrt{\langle
  v^2(T)\rangle}}{\sigma_{2-2}}\,, && \lambda_L  \sim \rho(T) \sqrt{\langle   v^2(T)\rangle}{\sigma_{2-2}} \, ,
\end{align}
with $\sigma_{2-2}$ the 2-to-2 scattering rate. As a final comment, we
note that the evolution of the unstable hydrodynamic mode, albeit
driven to instability with a choice of an imaginary momentum, may not
only grow with exponential growth faster than $\text{Im}[\omega] >
\lambda_L = 2\pi T$ but can also have a local group velocity larger
than $v_B$ at various values of imaginary $k$
(cf. Fig. \ref{fig:qnms}). As also found in
\cite{Hartman:2017hhp,Lucas:2017ibu}, this indicates that the butterfly
velocity may not in all generality be a bounding velocity. Understanding the relation between these observations and bounds on $\lambda_L$ and the speed of propagation of quantum correlations remains an important open problem, as does a better understanding of the relation between many-body microscopic chaos and instability-induced collective hydrodynamic turbulence. 

{\bf Acknowledgements.---}We are grateful to Mike Blake, Luis Lehner, Hong Liu, Andy Lucas, Subir Sachdev, Wilke van der Schee and Andrei Starinets for useful discussions. This research was supported in part by a VICI award of the Netherlands Organization for Scientific Research (NWO), by the Netherlands Organization for Scientific Research/Ministry of Science and Education (NWO/OCW), and by the Foundation for Research into Fundamental Matter (FOM). S. G. is also supported by the U.S. Department of Energy under grant Contract Number DE-SC0011090.

\bibliographystyle{apsrev4-1}
\bibliography{references}

\end{document}